\documentclass{elsarticle}
\makeatletter
\def\ps@pprintTitle{%
 \let\@oddhead\@empty
 \let\@evenhead\@empty
 \def\@oddfoot{}%
 \let\@evenfoot\@oddfoot}

\usepackage{lineno,hyperref}
\biboptions{numbers,sort&compress}
\modulolinenumbers[5]

\journal{arXiv.org}

\usepackage{enumitem}

\begin{document}

\begin{frontmatter}

\title{Class attendance, peer similarity, and academic performance in a large field study}


\author[address1]{Valentin Kassarnig\corref{corauth}}
\ead{kassarnig@ist.tugraz.at}
\author[address2,address3]{Andreas Bjerre-Nielsen}
\author[address4]{Enys Mones}
\author[address2,address4,address5]{Sune Lehmann}
\author[address2,address3]{David Dreyer Lassen}

\cortext[corauth]{Corresponding author. Institute of Software Technology, Graz University of Technology, Inffeldgasse 16B/II, 8010 Graz, Austria.}

\address[address1]{Institute of Software Technology, Graz University of Technology, Graz, Austria}
\address[address2]{Center for Social Data Science, University of Copenhagen, Copenhagen, Denmark}
\address[address3]{Department of Economics, University of Copenhagen, Copenhagen, Denmark}
\address[address4]{Department of Applied Mathematics and Computer Science, Technical University of Denmark, Kgs. Lyngby, Denmark}
\address[address5]{The Niels Bohr Institute, University of Copenhagen, Copenhagen, Denmark}

\begin{abstract}
Identifying the factors that determine academic performance is an essential part of educational research.
Existing research indicates that class attendance is a useful predictor of subsequent course achievements.
The majority of the literature is, however, based on surveys and self-reports, methods which have well-known systematic biases that lead to limitations on conclusions and generalizability as well as being costly to implement.
Here we propose a novel method for measuring class attendance that overcomes these limitations by using location and bluetooth data collected from smartphone sensors.
Based on measured attendance data of nearly 1,000 undergraduate students, we demonstrate that early and consistent class attendance strongly correlates with academic performance. 
In addition, our novel dataset allows us to determine that attendance among social peers was substantially correlated ($>$0.5), suggesting either an important peer effect or homophily with respect to attendance.\end{abstract}

\end{frontmatter}

\section*{Introduction}
An increasing number of individuals seek a high level of education to secure their future and improve their economic possibilities~\cite{carnevale2011college}.
Academic performance is an essential factor in the success of the post-education period with respect to employment~\cite{wise1975academic}. For this reason, the ability to predict students' academic success has been the subject of increasing interest. The knowledge regarding expected academic performance is also a valuable input for educators and school administrators, as this information can be used to identify and target vulnerable students at risk of dropping out or in need of additional attention.
However, gathering information about attendance levels using conventional methods (surveys or self-reports) is subject to inherent biases~\cite{eagle2009inferring} and moreover, can be costly to gather at the scale of schools or universities.

Here we propose a new method for measuring attendance. 
This new methodology overcomes important limitations of previous approaches. 
Specifically, our method leverages data collected via smartphone sensors to identify class locations from clusters of students following the same courses and estimate then the students' attendance (see Methods for details). 
We used the measured attendance levels of almost 1,000 university students to investigate the relationship between students' attendance and their grades, as well as the social aspects of academic performance. 
This is the first time a dataset of comparable richness has been used to conduct analyses on attendance.

The theoretical literature on student achievement emphasizes that class attendance is associated with better performance. One strand of theoretical literature are the pedagogical models where class attendance can be seen as student involvement, among other features which also highlights the resources of the school and the content being taught \cite{astin1984student}.
Other theoretical approaches include economic models where rational agents decide on optimal usage of time spent studying vs. leisure/other courses \cite{kelley1975student}. 

There is a large body of existing data-driven research on class attendance, absenteeism and their impact on academic achievements~\cite{schmidt1983maximizes,buckalew1986relationship,brocato1989much,park1990determinants,van1992class,romer1993students,durden1995effects,devadoss1996evaluation,gump2005cost,krohn2005student,lin2006cumulative,marburger2006does,stanca2006effects,chen2008class,crede2010class,nyamapfene2010does,westerman2011relationship,wang2014studentlife} as well as on the relationship between behavior of peers~\cite{card2013peer,ejrnaes2014should,winkelmann1999wages,hesselius2009sick,de2010absenteeism}.
However, the methodology applied in the large majority of previous work has limitations: results are based on analyzing surveys, sign-in-sheets or other types of self-reports, which are known to be prone to biases and errors~\cite{eagle2009inferring}.
Two exceptions that collected data from sensors are the \emph{StudentLife} study~\cite{wang2014studentlife} and Zhou et al.~\cite{zhou2016edum}. 
The \emph{StudentLife} study used location data recorded on students' phones~\cite{wang2014studentlife}.
Students were considered to be at class when they spent at least 90\% of the scheduled period at the class location. Although there were variations observed in class attendance, they found no relation between final grades and absence (either initial level or the pace of absence over the term) which contradicts the findings of most related studies. A likely explanation of the observations in the \emph{StudentLife} study is the small sample size ($<$ 50 students). 
Another approach is used by Zhou et al.~\cite{zhou2016edum} who  employ connectivity data from the WiFi network at Tsinghua University. The location of students and consequently their class attendance was determined by studying how students' phones connected to the nearly 2,800 WiFi access points with known locations distributed over university campus. Based on nine weeks of observations, Zhou et al. found that students with higher GPAs attended classes at a higher rate. Moreover, compared to low performing students, they were also found to be more likely to arrive late to class. 
Our approach shares some similarities with \cite{wang2014studentlife,zhou2016edum} but there are also some fundamental differences as we have also investigated dynamic patterns  (i.e. early and consistent attendance throughout the semester) and considered the social environment.  

Most existing studies have found that class attendance is a significant and positive predictor of course grades~\cite{schmidt1983maximizes,buckalew1986relationship,brocato1989much,park1990determinants,van1992class,romer1993students,durden1995effects,devadoss1996evaluation,gump2005cost,krohn2005student,lin2006cumulative,marburger2006does,stanca2006effects,chen2008class,crede2010class,westerman2011relationship}.
More specifically, the meta study by Crede et al.~\cite{crede2010class} concludes that attendance is the most accurate known predictor of academic performance, superior to scores on standardized admissions tests such as the SAT, high school GPA, study habits, and study skills.
Some studies report on experiments which have quantified the importance of attendance on exam performance through mandatory attendance~\cite{marburger2006does} and intentionally omitting parts of the curriculum~\cite{chen2008class}, and both found a significant effect on the number of exam questions answered correctly.
In addition to general attendance throughout the semester, initial attendance has been shown to be an important predictor of academic success~\cite{buckalew1986relationship}.
Previous results also indicate that average attendance drops over the course of the semester, irrespective of performance~\cite{van1992class,nyamapfene2010does,wangsmartgpa}.

The connection between attendance and peer behavior has been explored to a lesser extent.
Only a few preliminary observations exist in the literature based on co-occurrence at class or workplace~\cite{card2013peer,ejrnaes2014should,winkelmann1999wages,hesselius2009sick,de2010absenteeism}. One example is the work of De Paola, where individual absence behavior was found be related to peer group absenteeism~\cite{de2010absenteeism}. Yet these studies are limited by having no access to data regarding actual communication or interactions.

Educational policies aimed at increasing attendance have a broad set of tools. The standard approach is to enact mandatory attendance as experimented in~\cite{marburger2006does}. Other tools include tutoring \cite{roderick2014preventable} and nudging \cite{rogers2017randomized}. Moreover, for children and adolescents other options are to involve partnerships from schools to parents and their communities \cite{epstein2002present,sheldon2007improving}.

The aim of our study was to evaluate the accuracy of measuring class attendance from smartphone data and assess its usefulness for discovering new patterns in attendance. These aims were divided into three specific objectives:
\begin{enumerate}[label=(\roman*),itemsep=0ex,parsep=1ex]
\item To what extent could students' class attendance be inferred from data collected by smartphones? As a specific question we investigated whether a method based on finding large clusters of students enrolled in the same course was sufficient.
\item How accurate was the measure of students' attendance at predicting their subsequent exam performance?
\item What is the degree of similarity in attendance between students who are social peers?
\end{enumerate}

\section*{Materials and methods}
\subsection*{Data}
We employed data collected in the Copenhagen Networks Study (CNS)~\cite{measuringlargescale}.
As part of the CNS project, various data types were recorded by dedicated smartphones from nearly 1,000 undergraduate students at the Technical University of Denmark (DTU), over a period of 2 years. 
The channels for data collection include location (using GPS), proximity of other students (via Bluetooth scans) and mobile phone communication.
The CNS covered the academic years 2013/14 and 2014/15 which form the basis of our analysis.
About 78\% of the sampled students were male and 22\% female. They were divided in 24 different study lines (majors) and more than 60\% of them were newly enrolled in 2013; more than 25\% in 2012; the remaining enrolled in 2011. 
Their course grades and schedules were provided by the Technical University of Denmark. 

At the university, course attendance is non-mandatory and is not logged for the offered courses.
The educational model used in classes at the Technical University of Denmark is quite varied, ranging from lectures coupled with classroom exercises to more modern forms, e.g. flipped class room teaching or characterized by problem-based group work.
Also note that due to the possibility of exiting the experiment at any given point, the number of participants varied over time.

In terms of privacy all participating students have provided full consent for use of their data for research purposes. At any time a student could exit the study and request to have their data deleted. The collected data has been authorized by Danish Data Protection Agency. For further details of the CNS project and a short overview of the obtained data please refer to~\cite{measuringlargescale}.

\subsection*{Class location and attendance}
\begin{figure}[!ht]
	\includegraphics[width=\linewidth]{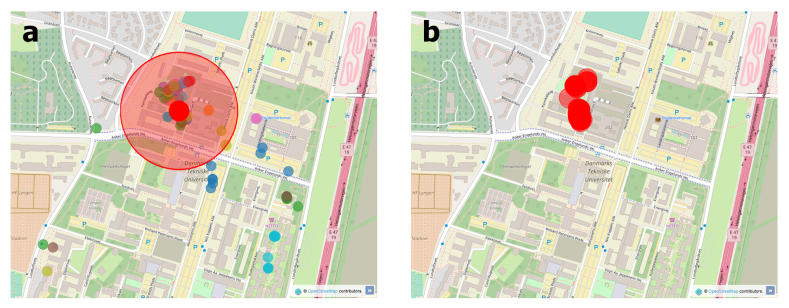}
	\caption{\textbf{Estimation of class locations at the Technical University of Denmark campus.}
		a) Location of the students who were assigned to a specific class (colored small circles), with the estimated location (moderate size red circle) and range of the class (large light red circle) in a particular day.
		b) Estimated class locations (red circles) of the same course throughout the semester.
		(Map data copyrighted OpenStreetMap contributors and available from \href{https://www.openstreetmap.org}{www.openstreetmap.org} under \href{https://creativecommons.org/licenses/by-sa/2.0/}{CC BY-SA 2.0})
	}
	\label{fig:class-location-viz}
\end{figure}

We first needed to calculate the locations of the classes before we could use them to determine the attendance of each eligible student (see Fig~\ref{fig:class-location-viz}).
The list of course participants and the associated class times were retrieved from the list of course grades and the course page archive, respectively.
To improve the accuracy of class estimation, we only considered courses with the standard DTU length of 4 hours and with at least 8 participants.
A significant fraction of classes did not have a fixed location throughout the 4-hour period as the students may change building or room (e.g. from a lecture hall to a laboratory).
Therefore, the designated time of the class was divided into sixteen 15-minute bins and for each time bin a separate class location was estimated, with our results being robust to changes of bin size.
For a particular time bin, we determined the proximity network of students from the Bluetooth scans representing nearby co-students within a distance of 15\,m.
In this network we identified the primary cluster, represented by the highest degree node (that is, the student surrounded by the most co-students signed up for the same course) and its direct neighbors.
In contrast to other procedures of determining the primary cluster (e.g. using the largest connected component), this method is robust against the noisy proximity data because some missing links do not necessarily affect the cluster.
Once the main cluster was found, the location of the class was defined by the median location of the members in the cluster, using his/her location with the highest accuracy during the 15-minute period.
If the estimated location was inside the university campus, we accept it as the location for the class in the corresponding time bin (shown in Fig~\ref{fig:class-location-viz}a).
The fact that our method uses Bluetooth data to identify class locations is a significant advantage, since we do not need to rely on official records. 
The lack of reliance on official records makes our approach applicable even when such records are not available or when the records do not match the actual class locations. 
This sensor based approach has (to the best of our knowledge) not been used previously.

The attendance of a participant was based on their location relative to the estimated class location in each bin.
All students who were no further than 200\,m from the estimated class location were assigned to the class in the specific time bin (all results were robust against variations in the distance threshold in the range 5 - 500\,m).
The value of 200\,m was explained by the noise in the measurement of location, especially when using GPS data inside buildings~\cite{moen1997accuracy}.
Members of the main cluster were automatically assigned to the class.
Fig~\ref{fig:class-location-viz}b shows the estimated locations for a specific course throughout the semester.
For further analyses we only considered students as attending when they were within the 200\,m range in at least three time bins to avoid false positives due to accidental proximity to the class location.

\begin{figure}[!ht]
	\includegraphics[width=\linewidth]{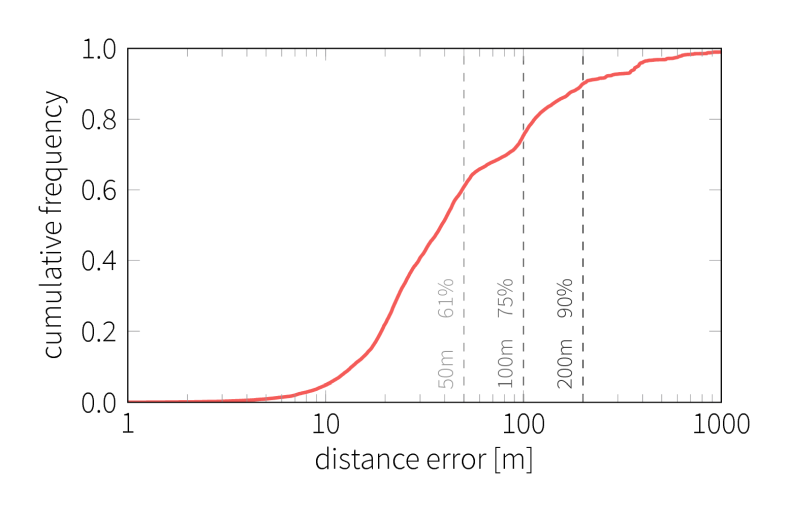}
	\caption{\textbf{Accuracy of class location inference.}
		The curve shows the cumulative distribution of distance error, that is, the distance between the designated location of the class and our estimation based on the location of the students.
		Dashed lines mark the distance errors of 50\,m, 100\,m and 200\,m along with the corresponding percentage of classes with error below those thresholds.
	}
	\label{fig:class-location-accuracy}
\end{figure}

We also tested our method against the actual course schedules and their locations.
Fig~\ref{fig:class-location-accuracy} shows the cumulative distribution of error in distance based on more than 26,000 class location estimations.
More than 75\% of estimated locations were found to be within the range of 100\,m of the scheduled location, and a 200\,m threshold includes 90\% of the classes.
Note that the error was estimated using the center of the corresponding building instead of the actual room which caused some imprecision for larger buildings.
Furthermore, there were cases of class relocations which did not appear in the official records and therefore the actual error was typically lower than our estimate. 

We have implemented an interactive visualization tool accompanying the paper that helps understand our approach. The tool and its source code is available on GitHub: \href{https://valentin012.github.io/class-loc-d3/}{ valentin012.github.io/class-loc-d3/}.

\subsection*{Social ties}
The exchange of text messages between two people suggests a strong social tie~\cite{van2010ll} and thus, we used this to infer the students' social connections.
That is, for each student, the list of their peers was constructed from those participants they had sent a message to or received a message from.
For the sake of simplicity and based on the sparsity of the network of text messages, we did not set any lower limit on the number of messages exchanged, that is, a single message is sufficient to establish a link in our network.

\subsection*{Statistical methods}
In order to measure the correlation between two observed variables we used Spearman's correlation coefficient. This nonparametric procedure does not assume a linear relation between the two variables since it only tests the association between their ranks instead of their raw values. The coefficient ranges from $+1$ to $-1$ indicating a perfect positive or negative correlation, respectively, whereas $0$ indicates no correlation. 

When comparing the distribution of observed variables from different groups we relied on the Kruskal--Wallis H-test. This nonparametric test examines whether all the observations originate from the same underlying population. As follow-up post-hoc test we used Dunn's multiple comparison test with Bonferroni correction to reveal which groups are significantly different from each other. 

In order to observe temporal trends in the data we used a Theil--Sen estimator. This non-parametric line-fitting technique is more robust against outliers and skewed data than, for instance, simple linear regression.

\section*{Results}
In the following we analyze attendance patterns for a group of students at the Technical University of Denmark (see Methods for an explanation of how attendance is computed).
First, we show that attendance is correlated with the achieved grades both at the level of a specific course and overall performance (i.e. average term grade). 
Second, we look at the temporal aspects and show that there is a general decrease in attendance over the course of a semester regardless of the performance.
However, the attendance behavior of low and high performing students displays substantial differences with respect to time. 
Finally, we show to what extent individual students share similar attendance patterns with their social peers. 

\subsection*{Academic achievement} 
\begin{figure}[!ht]
	\includegraphics[width=\textwidth]{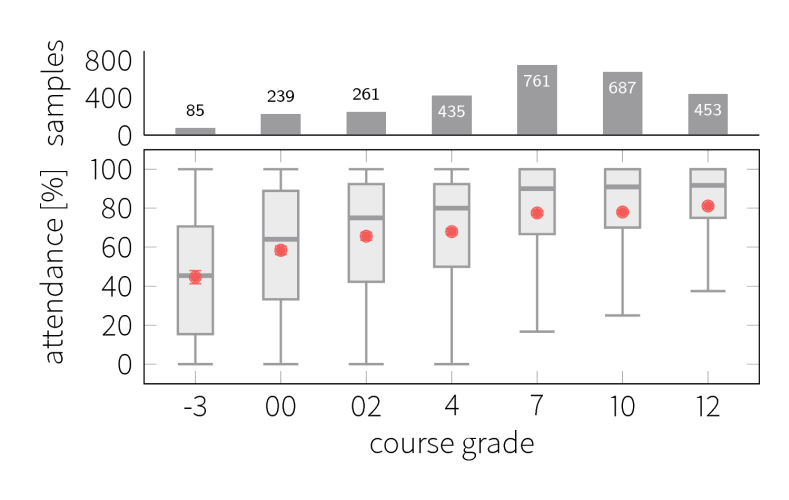}
	\caption{\textbf{Class attendance conditional on grade obtained.}
		Box plots show median values (solid horizontal line), mean values (red dots), lower and upper quartiles (box outline) and lower and upper fences (quartiles $\pm$ IQR, whiskers).
		Error bars mark standard deviation of the mean.
		Bar chart above the boxplot shows the number of observations in each grade group.}
	\label{fig:attendance-vs-grade}
\end{figure}

Fig~\ref{fig:attendance-vs-grade} shows the aggregate statistics for all grades (all courses and students considered) as a function of the final grade.
Grades follow the Danish grading system that spans between -3 and 12 with 7 distinct grade points (we also denote the corresponding grades according to the US system): -3 and 00 (grade F), 02 (grade D), 4 and 7 (grade C), 10 (grade B) and 12 (grade A).
A positive correlation can be observed in Fig~\ref{fig:attendance-vs-grade} with respect to the mean and median, which is quantified by a Spearman correlation of .255 ($p<.001$); the value of the coefficient indicates a weak ($<0.3$) positive correlation.
Observations where the individual fails to show up at the exam were excluded. 
The two plots at the boundary of the range (-3 and 12) show distinct, well-separated boxes marking a significant difference in the distributions.

\begin{figure}[!ht]
	\includegraphics[width=\textwidth]{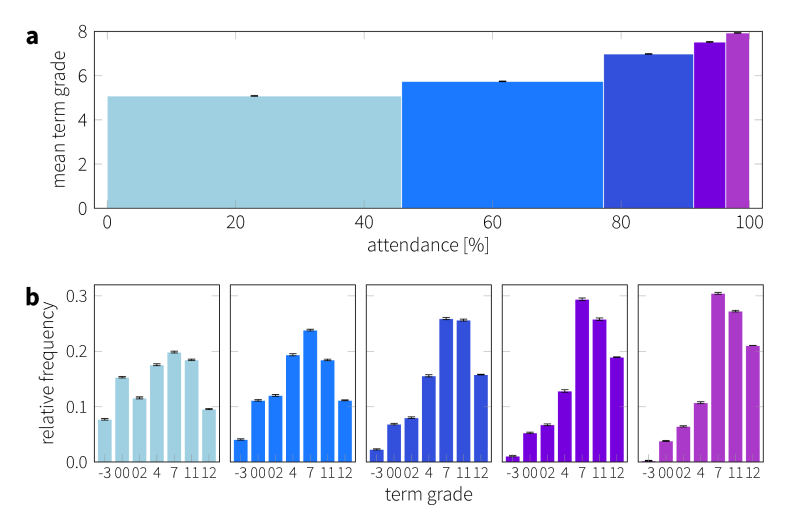}
	\caption{\textbf{Class attendance and performance.}
		a) Illustration of the five groups based on class attendance.
		Bars indicate groups of equal size, width corresponds to the span of attendance percentage in the specific group and height shows the mean term GPA.
		b) Grade distribution inside each attendance quintile.
		Each group includes at least 373 students.}
	\label{fig:attendance-groups}
\end{figure}

To further quantify the observed trend corresponding to the level of attendance, we divided the population into five quintiles based on the students' attendance and measure the performance inside these groups (see Fig~\ref{fig:attendance-groups}).
The distribution of attendance is illustrated in Fig~\ref{fig:attendance-groups}a, where each group is represented by a single bar and the width of the bars is adjusted to span over the covered attendance level.
The majority of the students were characterized by a high attendance (60\% of the students attend more than 75\% of the classes).
The height of the bars (mean term grade) show the observed correlation between attendance and performance.
The actual distribution of grades also shows variations over the attendance groups, as displayed in Fig~\ref{fig:attendance-groups}b.
Low attendance (leftmost) results in a broad distribution of grades, indicating that the performance is not solely a function of the attendance but strongly depends on other factors too.
However, groups of high attendance (middle to right) develop a peak at grade 7 and were characterized by an increasingly dominating likelihood of high grades.
The rate of failing (grade -3 and 00) in exams drops from 23\% (leftmost histogram) to less than 4\% (rightmost histogram).
Also note that there is no observable difference between the grade distribution of the last two attendance groups. The lack of difference suggests that attendance is better at discriminating between whether or not a student is likely to fail rather than predicting the actual grade achieved provided that the student passes the exam; this observation is supported by separate work on the CNS dataset~\cite{kassarnig2017predicting}, where the predictive power of not only class attendance, but of many other behavioral factors, is considered.
To statistically evaluate the variation in the distribution over the groups, we performed a Kruskal--Wallis H-test. This test rejected the global null hypothesis with $p<.001$ that the medians of the groups are all equal. A follow-up Dunn multiple comparison test with Bonferroni correction revealed pair-wise differences among the groups. The recorded $p$-values can be seen in Table~\ref{tab:attendance-grades-dist}.
All groups separated by at least one quintile were significantly different ($p<.001$), whereas the difference between some neighboring groups (e.g. Mod. vs. M-H, M-H vs. High) could be only confirmed at a lower significance level, further supporting our remarks in Fig~\ref{fig:attendance-groups}b.
Although attendance accounts for a significant fraction of variation observed in the distribution of grades, it should be noted that this does not necessarily indicate a causal effect.

\begin{table}[h!]
	\centering
	\begin{tabular}{lccccc}
		\hline
		& Low att. & L-M att. & Mod att. & M-H att. & High att. \\ \hline
		Low att.  & -        &          &          &          &           \\
		L-M att.  & .230     & -        &          &          &           \\
		Mod att.  & $<$.001  & $<$.001  & -        &          &           \\
		M-H att.  & $<$.001  & $<$.001  & .663     & -        &           \\
		High att. & $<$.001  & $<$.001  & $<$.001  & 1.0     & -         \\ \hline
	\end{tabular}
	\smallskip      
	\caption{Results of Dunn's multiple comparison test with Bonferroni correction for the grade distributions of different attendance groups. The table contains corrected $p$-values for each pairwise comparison, corresponding to the null hypothesis that the pair of groups has equal medians.}
	\label{tab:attendance-grades-dist}
\end{table}

\subsection*{Temporal effects} 
\begin{figure}[!ht]
	\includegraphics[width=\textwidth]{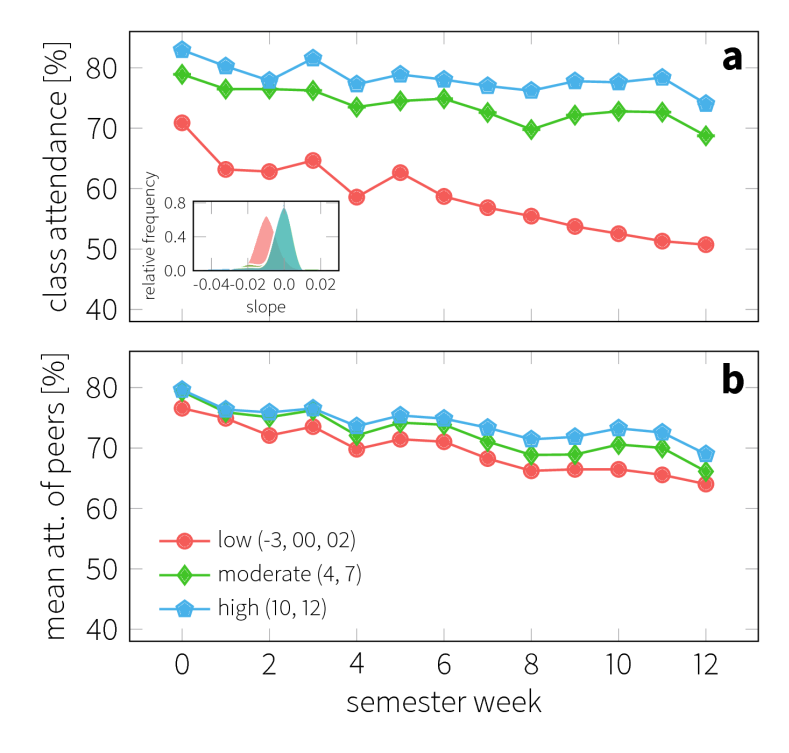}
	\caption{\textbf{Change in class attendance over a single semester.}
		a) Trends of attendance observed in the three performer groups: low (red circles), moderate (green diamonds) and high performers (blue pentagons), according to the Danish grading system.
		Inset shows the distribution of slopes measured for each pairs of data point in the trends.
		b) Mean attendance measured among the contacts of the students based on exchanged text messages.}
	\label{fig:attendance-trend}
\end{figure}

Besides differences in attendance across the population, there is also varying attendance for individual students over the duration of the semester (Fig~\ref{fig:attendance-trend}).
For the sake of simplicity, we divided our observation of student and course into three groups: the first group is characterized by low grades in the course (grades -3, 00 and 02); the second is moderate performance (grades 4 and 7,) and; finally high achievers (grades 10 and 12). 
With about 41\%, moderate performers constituted the largest fraction of nearly 8,400 observations, followed by high (37\%) and the low achievers (22\%).
We computed the average attendance for each group over the semester (see Fig~\ref{fig:attendance-trend}a) and observed a general decrease. Further, low performers showed a drop already in the first week with an attendance level 10\% lower than that of the high performers.
This initial difference increases further throughout the semester as the rate of absence among low performers is consistently higher compared to the moderate and high performer groups. 
The total drop in the attendance among low performers is above 20\% points, compared to the 10\% points and 8\% points observed among the moderate and high performers.
The corresponding Kruskal--Wallis H-tests rejected the global null hypothesis ($p<.001$) that the medians of the groups are all equal. The corresponding Dunn multiple comparison test with Bonferroni correction suggested significant difference between the observations from every pair of performance groups ($p<.05$ between mod. and high performers; $p<.001$ for others).
These differences in the trends were further supported by a Theil--Sen estimation for slopes: -1.4 \% points/week for low performers, opposed to the -.6 and -.4\% points/week measured in the moderate and high performer groups.
This difference in the slopes is portrayed in the accompanying inset of Fig~\ref{fig:attendance-trend}a, where we show the distribution of slopes measured in all pairs of data points in the trends of the main plot.
The low performers' distribution of slopes is clearly separated and significantly different from those describing moderate and high performers ($p<.001$ for low vs. mod. and $p<.01$ for low vs. high).
Note that differences in the decrease in attendance are not significant between moderate and high performers ($p=1.0$), also supporting the hypothesis that (the absence of) attendance affects the failing rate to a higher extent than the actual grades.
Finally, at the end of the semester, the initial difference inflates to 24\% points between low and high performers due to the faster dropping rate.
In summary, the temporal attendance behavior of different types of students differs in the way that low performers start out with a lower attendance rate which also decreases more rapidly throughout the semester compared to that of moderate and  high performers. 

\subsection*{Peer similarity}

Next, we investigated the social aspects of academic performance and attendance: the plot in Fig~\ref{fig:attendance-trend}b illustrates the mean attendance level of the peers in the different performance groups.
For each individual we created a list of their strongest ties and then calculated the average attendance of them.
Strong social contacts were inferred from text message exchanged between two students, as this form of communication indicates a strong bond~\cite{van2010ll}.
On average, each student has exchanged text messages with 4.4 other students from among the participants.
Surprisingly, we observed the same differences as above, although less pronounced than in the individual trends: the peers of the low performers also display lower attendance. 
This supports previous findings on homophily and peer effects: students communicate more with others who are similar in performance (note that contacts are based on text messages and not on physical proximity).

\begin{figure}[!ht]
	\includegraphics[width=\textwidth]{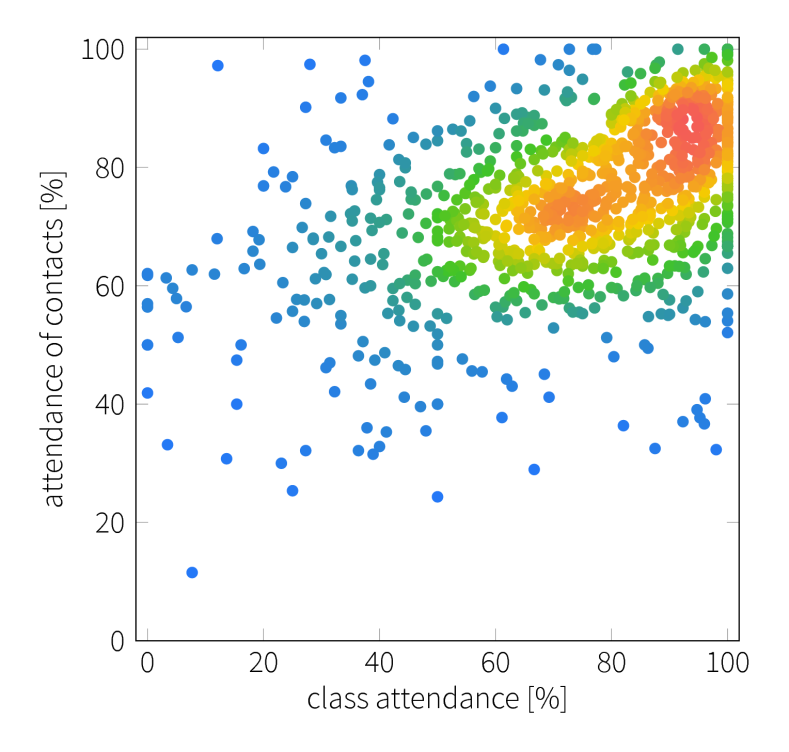}
	\caption{\textbf{Correlation between own and peers' attendance.}
		Scatter plot that shows student's own attendance vs mean attendance among contacts (inferred from text messages) at the course level.
		Color represents the relative density of data points.}
	\label{fig:attendance-friends}
\end{figure}

\begin{figure}[!ht]
	\includegraphics[width=\textwidth]{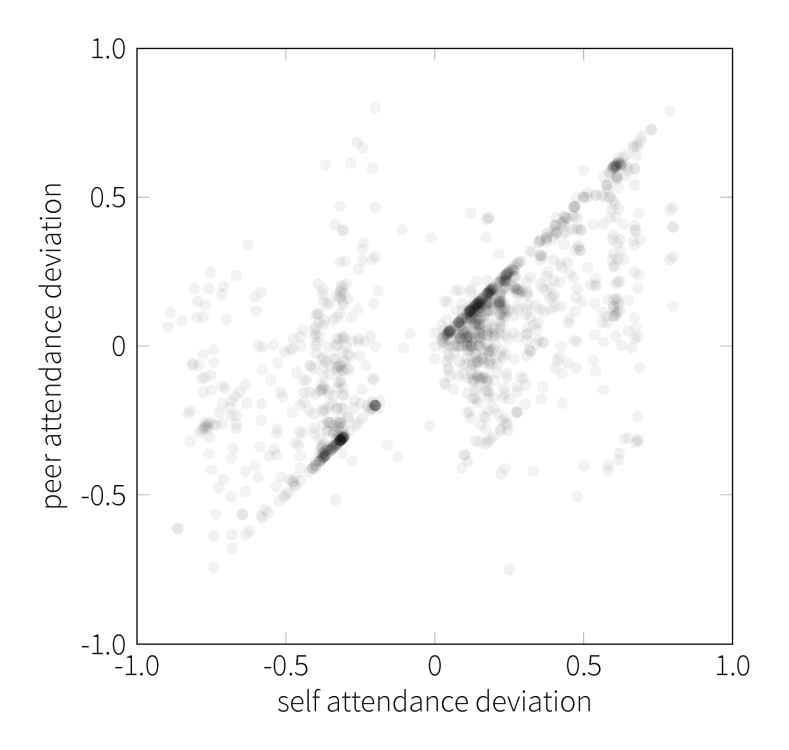}
	\caption{\textbf{Correlation between self and peer corrected attendance.}
		Shown are the attendance relative to the average calculated at the level of classes.
		Each dot represents a student's own and their peers' median deviation from the average attendance with respect to a course, that is, their attendance deviation averaged over all classes of a course.}
	\label{fig:attendance-corrected}
\end{figure}

The observed correlation between own attendance and attendance measured in the ego-networks (the student and their contacts) is clear at the individual level as well.
Fig~\ref{fig:attendance-friends} shows the attendance of contacts as a function of the students own attendance, along with the density of observations (color of the dots in plot).
Similarity between own and peers' attendance is visible in this scatter plot and has a moderate correlation of .48 ($p<.001$).
Furthermore, as shown by the relative density of data points in Fig~\ref{fig:attendance-friends}, the number of students is characterized by a peak in peer attendance for high own attendance (above 60\%).
In other words, peer attendance has a narrow distribution at high attendance levels, compared to the more broad distribution observed below own attendance of 50\%.
The pattern of similarity is robust against removal of class-level effects as seen in Fig~\ref{fig:attendance-corrected} of the Supporting Information.
The class effects were removed by subtracting the mean attendance for each class that took place. Robustness of the results suggests that the similarity is driven by homophily or peer effects.

\section*{Discussion}
In this paper, we introduced a novel high precision method to measure class attendance in an academic setting.
Contrary to previous studies, our method overcomes various limitations caused by restricted data collection techniques.
The accuracy of the method makes it possible to scale our measurement and allows for very high precision inference of attendance compared to standard survey-based measurements.
Applying our method to a population of nearly 1,000 undergraduate students, we have shown that in this population attendance is not only weakly correlated ($<0.3$) with academic success (supporting previous studies) but it is also reflected in the social interactions, which show that students of similar performance tend to be clustered in the network.
Our results suggest a strong mixture of either sorting or peer effects which appears early in the semester and spans across the entire term. 

Based on the detailed measurements, we then investigated the temporal aspects of attendance.
A general decreasing trend describes the entire population in our dataset, however, there is a clear difference with respect to the pace and level of the effect conditional on performance, indicating a strong early differentiation in behavior between low performers and the rest of the population.
Interestingly, this effect vanishes when moderate and high performers are compared.
This distinction between low performers and all others is also present in the aggregate statistics of attendance: rate of failing drops by 80\% (compared to the value measured in the low attendance quintile) once the attendance reaches the 75\% threshold.
These results indicate that the effect of attendance on performance shows a complex pattern: while attendance is a strong predictive measure for the failing rate, the effect is less pronounced at high attendance levels and high performance.

Using the contacts obtained from mobile phone communications, we were able to investigate the social aspects of attendance.
We found that homophily is present in all groups of academic performance levels, however, it is stronger among high performers.
This is supported by an overall robust trend observed in the relative attendance (compared to peers) among high performers, as well as in the correlation between own and peer attendance at the individual level.

We note, however, that our method and the results have some limitations that we address in the following.
First, estimation of class locations is based on Bluetooth and GPS signals, both of which are subject to noise.
We briefly outline how we mitigate these errors.
Our approach is to first identify clusters of physical meetings.
We do this by including every Bluetooth signal found between two devices, irrespective of the signal strength, which results in contacts within a typical distance of 10-15\,m~\cite{sekara2014thestrength}.
Although scans may fail, we reduce the resulting error by employing scans from both devices between two students -- this only requires one successful scan and therefore the error should be minimal. 
Subsequently we employ the identified clusters to estimate the class location.
By using the median location of the most connected student within the cluster, we remove noise inherent in location data.
After applying our corrections, less than 10\% of the estimated class locations are found to be outside of the 200\,m radius around the official locations.

Note that it is beyond the scope of this study to control for personal characteristics.
However, in~\cite{kassarnig2017predicting}, attendance is shown to be an important predictor of subsequent performance when controlling for performance of friends and personality.
Also, our estimates are not causal estimates of attendance, cf.~\cite{marburger2006does,chen2008class}, as the choice of attendance is likely to be correlated with other factors.
That is, when a student attends more classes it is not likely to lead to a change in achievements equal to the one we observe.
This is due to the fact that attendance and performance could be driven by same latent factor which is both fixed and unobserved to us as researchers.

Another limitation comes from the fact that we have measurements on a subset of all students enrolled in the courses.
As an illustration, around 40\% of first year students accepted a phone from among the entire cohort of freshmen.
The students who participated in the study are different from the average student as they achieve higher grades~\cite{bjerre2017dynamics}.
We nevertheless observe high variation with respect to attendance and performance within our dataset and, thus, we believe that our results are appropriate precursors of the trends present in the larger student population.

We have demonstrated the connection between attendance and performance; however, attendance alone does not imply active participation.
Students who attend class may or may not participate actively in class activities, and although no significant correlation has been observed e.g. between seating position and performance~\cite{buckalew1986relationship}, it is still unknown what other factors contribute to the academic performance.
Our methods could be used in conjunction with further experiments such as~\cite{marburger2006does,chen2008class} to yield additional insights on effects of class attendance.

Finally, academic performance is a complex question with multiple facets, and limiting the measurement of success to the raw value of grades is an oversimplification of a high dimensional problem.
A first concern is that students were susceptible to different subjects and show interest in distinct fields.
A more detailed analysis of performance could, e.g., restrict the analysis above to single subjects.
Although this would provide a better understanding of individual performance, the aim of the paper was to investigate the connection at the basic level of attendance.
Results with different performance levels suggest that attendance is an efficient predictor of failing, indicating that differentiation at higher orders (that is, among good performers) indeed requires more detailed knowledge regarding the individuals themselves.
Further research is therefore needed to understand how factors beyond attendance influence academic success.

\section*{Supporting information}
\paragraph*{S1 File.}
\label{S1_File}
{\bf Dataset details.} The supporting information contains further details on the data collection process and the recorded data types. Additionally, we discuss our method on correcting attendance data on the class level and show that we observe distinct attendance-performance correlations for courses of different subject areas.

\section*{Acknowledgments}
The entire Copenhagen Networks Study, including the current study, has been approved by the Danish Data Protection Agency (DDPA), which is the relevant legal entity in Denmark.
The data used in this study includes third-party reported information on grades obtained from administrative sources. According to the Act on Processing of Personal Data, such data cannot be made available in the public domain.
We confirm that the data is available upon request to all interested researchers under conditions stipulated by the DDPA. Data inquiries should be addressed to the Social Fabric steering committee, to be reached at ddl@econ.ku.dk.

\nolinenumbers


\begin{thebibliography}{}
\bibitem{carnevale2011college}
Carnevale AP, Rose SJ, Cheah B.
\newblock The College Payoff: Education, Occupations, Lifetime Earnings.
\newblock Georgetown University Center on Education and the Workforce. 2011;.

\bibitem{wise1975academic}
Wise DA.
\newblock Academic achievement and job performance.
\newblock The American Economic Review. 1975;65(3):350--366.



\bibitem{eagle2009inferring}
Eagle N, Pentland AS, Lazer D.
\newblock Inferring friendship network structure by using mobile phone data.
\newblock Proceedings of the national academy of sciences.
2009;106(36):15274--15278.

\bibitem{astin1984student}
Astin AW.
\newblock Student Involvement: A Developmental Theory for Higher Education.
\newblock Journal of college student personnel. 1984; 25(4): 297--308.

\bibitem{kelley1975student}
Kelley AC.
\newblock The student as a utility maximizer.
\newblock The Journal of Economic Education. 1975;6(2): 82--92.



\bibitem {schmidt1983maximizes}
Schmidt RM.
\newblock Who maximizes what? A study in student time allocation.
\newblock The American Economic Review. 1983;73(2):23--28.


\bibitem{buckalew1986relationship}
Buckalew L, Daly JD, Coffield K.
\newblock Relationship of initial class attendance and seating location to
academic performance in psychology classes.
\newblock Bulletin of the Psychonomic Society. 1986;24(1):63--64.

\bibitem{brocato1989much}
Brocato J.
\newblock How much does coming to class matter? Some evidence of class
attendance and grade performance.
\newblock Educational Research Quarterly. 1989;.

\bibitem{park1990determinants}
Park KH, Kerr PM.
\newblock Determinants of academic performance: A multinomial logit approach.
\newblock The Journal of Economic Education. 1990;21(2):101--101.

\bibitem{van1992class}
Van~Blerkom ML.
\newblock Class attendance in undergraduate courses.
\newblock The Journal of psychology. 1992;126(5):487--494.

\bibitem{romer1993students}
Romer D.
\newblock Do students go to class? Should they?
\newblock The Journal of Economic Perspectives. 1993; 7(3):167--174.

\bibitem{durden1995effects}
Durden GC, Ellis LV.
\newblock The effects of attendance on student learning in principles of economics.
\newblock The American Economic Review. 1995; 85(2): 343--346




\bibitem{devadoss1996evaluation}
Devadoss S, Foltz J.
\newblock Evaluation of factors influencing student class attendance and
performance.
\newblock American Journal of Agricultural Economics. 1996;78(3):499--507.

\bibitem{gump2005cost}
Gump SE.
\newblock The cost of cutting class: Attendance as a predictor of success.
\newblock College Teaching. 2005;53(1):21--26.

\bibitem{krohn2005student}
Krohn GA, O'Connor CM.
\newblock Student effort and performance over the semester.
\newblock The Journal of Economic Education. 2005;36(1):3--28.

\bibitem{lin2006cumulative}
Lin TF, Chen J.
\newblock Cumulative class attendance and exam performance.
\newblock Applied Economics Letters. 2006;13(14):937--942.

\bibitem{marburger2006does}
Marburger DR.
\newblock Does mandatory attendance improve student performance?
\newblock The Journal of Economic Education. 2006;37(2):148--155.

\bibitem{stanca2006effects}
Stanca L.
\newblock The effects of attendance on academic performance: Panel data
evidence for introductory microeconomics.
\newblock The Journal of Economic Education. 2006;37(3):251--266.

\bibitem{chen2008class}
Chen J, Lin TF.
\newblock Class attendance and exam performance: A randomized experiment.
\newblock The Journal of Economic Education. 2008;39(3):213--227.

\bibitem{crede2010class}
Cred{\'e} M, Roch SG, Kieszczynka UM.
\newblock Class attendance in college a meta-analytic review of the
relationship of class attendance with grades and student characteristics.
\newblock Review of Educational Research. 2010;80(2):272--295.

\bibitem{nyamapfene2010does}
Nyamapfene A.
\newblock Does class attendance still matter?
\newblock engineering education. 2010;5(1):64--74.


\bibitem{westerman2011relationship}
Westerman JW, Perez-Batres LA, Coffey BS, Pouder RW.
\newblock The relationship between undergraduate attendance and performance
revisited: Alignment of student and instructor goals.
\newblock Decision Sciences Journal of Innovative Education. 2011;9(1):49--67.

\bibitem{wang2014studentlife}
Wang R, Chen F, Chen Z, Li T, Harari G, Tignor S, et~al.
\newblock Studentlife: assessing mental health, academic performance and
behavioral trends of college students using smartphones.
\newblock In: Proceedings of the 2014 ACM International Joint Conference on
Pervasive and Ubiquitous Computing. ACM; 2014. p. 3--14.


\bibitem{card2013peer}
Card D, Giuliano L.
\newblock Peer effects and multiple equilibria in the risky behavior of
friends.
\newblock Review of Economics and Statistics. 2013;95(4):1130--1149.

\bibitem{ejrnaes2014should}
Ejrn{\ae}s M, Holm A, Le~Maire D.
\newblock Should I Stay or Should I Go: Peer Effects in Absenteeism.
\newblock Centre for Applied Microeconometrics - University of Copenhagen;
2014. 3.




\bibitem{winkelmann1999wages}
Winkelmann R.
\newblock Wages, firm size and absenteeism.
\newblock Applied Economics Letters. 1999;6(6):337--341.

\bibitem{hesselius2009sick}
Hesselius P, Nilsson JP, Johansson P.
\newblock Sick of your colleagues' absence?
\newblock Journal of the European Economic Association. 2009;7(2-3):583--594.

\bibitem{de2010absenteeism}
De~Paola M.
\newblock Absenteeism and peer interaction effects: evidence from an Italian
public institute.
\newblock The Journal of Socio-Economics. 2010;39(3):420--428.


\bibitem{zhou2016edum}
Zhou M, Ma M, Zhang Y, SuiA K, Pei D, Moscibroda T.
\newblock EDUM: classroom education measurements via large-scale WiFi networks.
\newblock In: Proceedings of the 2016 ACM International Joint Conference on
Pervasive and Ubiquitous Computing. ACM; 2016. p. 316--327.

\bibitem{wangsmartgpa}
Wang R, Harari G, Hao P, Zhou X, Campbell AT.
\newblock SmartGPA: how smartphones can assess and predict academic performance
of college students.
\newblock In: Proceedings of the 2015 ACM International Joint Conference on
Pervasive and Ubiquitous Computing. ACM; 2015. p. 295--306.


\bibitem{roderick2014preventable}
Roderick M, Kelley-Kemple T, Johnson DW and Beechum NO.
\newblock Preventable Failure: Improvements in Long-Term Outcomes When High Schools Focused on the Ninth Grade Year.
\newblock University of Chicago Consortium on Chicago School Research; 2014.

\bibitem{rogers2017randomized}
Rogers T, Duncan T and Wolford T, Ternovski J, Subramanyam S and Reitano, A.
\newblock A Randomized Experiment Using Absenteeism Information to" Nudge" Attendance. 
\newblock Regional Educational Laboratory Mid-Atlantic, no. 252; 2017.

\bibitem{epstein2002present}
Epstein JL and Sheldon SB.
\newblock Present and accounted for: Improving student attendance through family and community involvement
\newblock The Journal of Educational Research. 2002;95(5):308--318.

\bibitem{sheldon2007improving}
Sheldon SB.
\newblock Improving student attendance with school, family, and community partnerships
\newblock The Journal of Educational Research. 2007;100(5):267--275.


\bibitem{measuringlargescale}
Stopczynski A, Sekara V, Sapiezynski P, Cuttone A, Madsen MM, Larsen JE, et~al.
\newblock Measuring large-scale social networks with high resolution.
\newblock PloS one. 2014;9(4):e95978.

\bibitem{moen1997accuracy}
Moen R, Pastor J, Cohen Y.
\newblock Accuracy of GPS telemetry collar locations with differential
correction.
\newblock The Journal of Wildlife Management. 1997; p. 530--539.

\bibitem{van2010ll}
Van~Cleemput K.
\newblock “I’ll see you on IM, text, or call you”: A social network
approach of adolescents’ use of communication media.
\newblock Bulletin of Science, Technology \& Society. 2010;30(2):75--85.

\bibitem{kassarnig2017predicting}
Kassarnig V, Mones E, Bjerre-Nielsen A, Sapiezynski P, Lassen DD, Lehmann S.
\newblock Academic Performance and Behavioral Patterns; 2017.
\newblock arXiv:1706.09245.

\bibitem{sekara2014thestrength}
Sekara V, Lehmann S.
\newblock The Strength of Friendship Ties in Proximity Sensor Data.
\newblock PLoS ONE. 2014;9(7):e100915.

\bibitem{bjerre2017dynamics}
Bjerre-Nielsen A, Dreyer~Lassen D.
\newblock Opportunity and Similarity in Dynamic Friendships; 2017.



\end{thebibliography}
\end{document}